\documentclass[english,aps, twocolumn, pre,superscriptaddress]{revtex4-1}
\usepackage[T1]{fontenc}
\usepackage[latin9]{inputenc}
\usepackage{color}
\usepackage{array}
\usepackage{booktabs}
\usepackage{multirow}
\usepackage{amsmath}
\usepackage{amssymb}
\usepackage{graphicx}
\makeatletter
\providecommand{\tabularnewline}{\\}
\usepackage[colorlinks=true,linkcolor=MidnightBlue,urlcolor=MidnightBlue,citecolor=MidnightBlue,anchorcolor=MidnightBlue]{hyperref}\usepackage[dvipsnames]{xcolor}
\makeatother
\usepackage{babel}
\begin{document}
\title{Fast Bayesian inference of optical trap stiffness and particle diffusion}
\author{Sudipta Bera}
\affiliation{Dept of Physical Sciences, Indian Institute of Science Education
and Research, Kolkata, Mohanpur 741246}
\author{Shuvojit Paul}
\affiliation{Dept of Physical Sciences, Indian Institute of Science Education
and Research, Kolkata, Mohanpur 741246}
\author{Rajesh Singh}
\affiliation{The Institute of Mathematical Sciences-HBNI, CIT Campus, Taramani,
Chennai 600113}
\author{Dipanjan Ghosh}
\affiliation{Dept of Chemical Engineering, Jadavpur University, Kolkata 700032}
\author{Avijit Kundu}
\affiliation{Dept of Physical Sciences, Indian Institute of Science Education
and Research, Kolkata, Mohanpur 741246}
\author{Ayan Banerjee}
\email{ayan@iiserkol.ac.in}
\affiliation{Dept of Physical Sciences, Indian Institute of Science Education
and Research, Kolkata, Mohanpur 741246}
\author{R. Adhikari}
\email{rjoy@imsc.res.in}
\affiliation{The Institute of Mathematical Sciences-HBNI, CIT Campus, Taramani,
Chennai 600113}
\begin{abstract}
Bayesian inference provides a principled way of estimating the parameters
of a stochastic process that is observed discretely in time. The overdamped
Brownian motion of a particle confined in an optical trap is generally
modelled by the Ornstein-Uhlenbeck process and can be observed directly
in experiment. Here we present Bayesian methods for inferring the
parameters of this process, the trap stiffness and the particle diffusion
coefficient, that use exact likelihoods and sufficient statistics
to arrive at simple expressions for the maximum a posteriori estimates.
This obviates the need for Monte Carlo sampling and yields methods
that are both fast and accurate. We apply these to experimental data
and demonstrate their advantage over commonly used non-Bayesian fitting
methods. 
\\\\DOI:
\href{http://dx.doi.org/10.1038/srep41638}{10.1038/srep41638}\\ 
\end{abstract}
\maketitle
\section{Introduction}
Since the seminal contributions of Rayleigh, Einstein, Smoluchowski,
Langevin and others \cite{chandrasekhar1943stochastic}, stochastic
processes have been used to model physical phenomena in which fluctuations
play an essential role. Examples include the Brownian motion of a
particle, the fluctuation of current in a resistor, and the radioactive
decay of subatomic particles \cite{van1992stochastic}. A central
problem is to infer the parameters of the process from partially observed
sample paths, for instance, the diffusion constant from a time series
of positions, or the resistance from a time series of current measurements,
and so on. Bayesian inference provides a principled solution to this
inverse problem \cite{jeffreys1998theory}, making optimal use of
the information contained in the partially observed sample path \cite{zellner1988optimal}. 

The motion of a Brownian particle harmonically trapped in optical
tweezers in a volume of a viscous fluid far away from walls is usually
modelled by the Ornstein-Uhlenbeck stochastic process \cite{van1992stochastic,gardiner1985handbook}.
The stiffness $k$ of the harmonic potential, the friction $\gamma$
of the particle, and the temperature $k_{B}T$ of the fluid are the
three parameters of the stochastic dynamics. For spherical particles
Stokes' law $\gamma=6\pi\eta a$ relates the friction to the particle
radius $a$ and the fluid viscosity $\eta$, while the Einstein relation,
which holds generally, relates the particle diffusion coefficient
$D$ to the temperature and friction through $D=k_{B}T\gamma^{-1}$
\cite{kubo1966fluctuation}. Of these several physical parameters,
any \emph{two} may be chosen independently, and it is conventional
to choose the ratio $k/\gamma$ and $D$ to be independent as they
relate, respectively, to the mean regression rate $\lambda$ and the
volatility $\sigma$ of the Ornstein-Uhlenbeck process (see below). 

Reliable estimation of the stiffness is a necessary first step in
using tweezers for force measurements. An estimation of the friction,
for a particle of known size, provides an indirect measure of the
viscosity of the medium. This microscopic method of viscometry is
of great utility when sample volumes are in the nanoliter range and
conventional viscometric methods cannot be used. Conversely, an estimate
of the friction in a fluid of known viscosity provides a method for
estimating the particle size. In both these cases, an estimate of
the diffusion coefficient provides, by virtue of the Einstein relation,
identical information.

Extant protocols for estimating these parameters from discrete observations
of the position of the Brownian particle can be divided into ``fluctuation''
and ``response'' categories. In the fluctuational methods, the fluctuating
position of the particle is recorded and the known forms of the static
and dynamic correlation functions are fitted to the data. In response
methods, external perturbations are applied to the particle and the
known forms of the average response is fitted to the data. Considerable
care is needed in these fitting procedures to obtain reliable estimates
\cite{berg2004power}. 

In recent work \cite{richly2013calibrating}, Bayesian inference has
been applied to the optical tweezer parameter estimation problem.
The posterior probability distribution of the stiffness and diffusion
coefficient is estimated for a time series of positions, making it
a method of the ``fluctuation'' category. Monte Carlo sampling is
needed to compute the posterior distribution and estimation from a
time series of $10,000$ points requires few tens of seconds. The
advantages of the Bayesian method over conventional calibration methods
have been discussed at length in this work.

In this paper, we present two Bayesian methods, of the fluctuational
category, which do not require Monte Carlo sampling and, consequently,
are extremely fast. For example, they estimate the trap stiffness
and diffusion coefficient from time series containing a million points
in less than a millisecond. The first method extracts information
exploiting the Markov property of the sample path and jointly estimates
the mean regression rate $k/\gamma$ and the diffusion coefficient
$D$. The second method extracts information from the equal-time fluctuations
of the position, which, in equilibrium, cannot depend on the friction
coefficient, and is, then, a function of the stiffness $k$ alone.
In essence, this is a recasting of the ``equipartition'' method
in the language of Bayesian inference.

The first method, in addition to inheriting the generic advantages
of Bayesian inference that have already been pointed out in \cite{richly2013calibrating},
has several specific advantages. First, it uses the exact expression
for the likelihood, which is valid for any $\Delta t$, the interval
at which the position is observed. Therefore, it works reliably with
data acquired at low frequencies. Second, the exact likelihood is
expressed in terms of four sufficient statistics, which are quadratic
functions of the positions. Their use greatly reduces the computation
needed to evaluate the posterior distribution, as four numbers, rather
than a large time series now represents the entire information relevant
to inference. Finally, we are able to obtain exact maximum a posteriori
(MAP) estimates of the mean regression and diffusion coefficients,
and their error bars, in terms of the four sufficient statistics.
This obviates the need for the Monte Carlo sampling or numerical minimization
steps usually required in Bayesian inference. Bayesian credible regions
are easily calculated from the analytically obtained error bars. The
second method is different from the conventional equipartition method
in that it provides a Bayesian error bar, representing a Bayesian
credible interval, rather than a frequentist confidence interval \cite{jaynes1976confidence}.
The combined use of exact likelihoods, sufficient statistics and analytical
MAP estimates yields both speed and accuracy in parameter estimation.

We apply both methods to experimental data and obtain MAP estimates
and error bars that are in excellent agreement with each other. These
estimates are found to be in good agreement with the commonly used
power spectral density calibration method \cite{berg2004power}. The
Bayesian methods of this paper are implemented in a well-documented,
open-source software freely available on GitHub \cite{pybisp}.

The remainder of the paper is organized as follows. In the next section
we recall several key properties of the sample paths and distributions
of the Ornstein-Uhlenbeck process. In Section \ref{sec:Bayesian-inference},
we present the Bayesian methods, in section \ref{sec:Experimental-setup-and}
we describe the experimental setup, and in section \ref{sec:Results}
we apply the Bayesian procedures to the experimental data. We conclude
with a discussion of future directions in the application of Bayesian
inference to optical tweezer experiments and advocate its use as a
complement to standard non-Bayesian methods.

\section{Ornstein-Uhlenbeck process\label{sec:Ornstein-Uhlenbeck-process}}

The Langevin equation for a Brownian particle confined in a potential
$U$ is given by 
\begin{equation}
m\dot{v}+\gamma v+\nabla U=\xi
\end{equation}
where $\xi(t)$ is a zero-mean Gaussian white noise with variance
$\langle\xi(t)\xi(t')\rangle=2k_{B}T\gamma\delta(t-t')$ as required
by the fluctuation-dissipation theorem. In the limit of vanishing
inertia and a harmonic potential, $U=\frac{{1}}{2}kx^{2}$, we obtain
the overdamped Langevin equation 
\begin{equation}
\dot{x}=-\frac{k}{\gamma}x+\sqrt{\frac{2k_{B}T}{\gamma}}\zeta(t),
\end{equation}
where $\zeta(t)$ is now a zero-mean Gaussian white noise with unit
variance. This describes the Ornstein-Uhlenbeck process, whose sample
paths obey the Ito stochastic differential equation 
\begin{equation}
dx=-\lambda xdt+\sigma dW,
\end{equation}
where $\lambda$ is the mean-regression rate, $\sigma$ is the volatility
and $W(t)$ is the Wiener process \cite{gardiner1985handbook}. 

For Brownian motion, the mean regression rate $\lambda=k/\gamma$
is the ratio of the stiffness and the friction while the square of
the volatility $\sigma^{2}=2D$ is twice the diffusion coefficient
$D$. The latter follows by comparing the Langevin and Ito forms of
the path equation and recalling the Einstein relation $D=k_{B}T\gamma^{-1}$
between the diffusion and friction coefficients of a Brownian particle.
In problems involving Brownian motion, it is convenient to work with
the diffusion coefficient, rather than the volatility. 

The ratio of $\lambda$ and $D$ provides the stiffness 
\begin{equation}
\lambda/D=k/k_{B}T
\end{equation}
in units of $k_{B}T$. We note that, in general, there is no relation
between the mean regression rate and volatility of the Ornstein-Uhlenbeck
process and the preceding identity is a consequence of additional
$physical$ constraints, namely the fluctuation-dissipation and Einstein
relations \cite{kubo1966fluctuation}. 

The transition probability density $P_{1|1}$$(x^{\prime}t^{\prime}|x\thinspace t$),
the probability of a displacement from $x$ at time $t$ to $x^{\prime}$
at time $t^{\prime}$, obeys the Fokker-Planck equation $\partial_{t}P_{1|1}=\mathcal{L}P_{1|1}$,
where the Fokker-Planck operator is 

\begin{equation}
\mathcal{L}=\frac{\partial}{\partial x}\lambda x+\frac{\partial^{2}}{\partial x^{2}}D.
\end{equation}
The solution is a normal distribution, 

\begin{equation}
x^{\prime}t^{\prime}|x\thinspace t\sim\mathcal{N}\left(xe^{-\lambda(t^{\prime}-t)},\frac{D}{\lambda}[1-e^{-2\lambda(t^{\prime}-t)}]\right),
\end{equation}
where $\mathcal{N}(a,b)$ is the univariate normal distribution with
mean $a$ and variance $b$. This solution is exact and holds for
arbitrary values of $|t-t^{\prime}|$. The correlation function decays
exponentially, 
\begin{equation}
C(t-t^{\prime})\equiv\langle x(t)x(t^{\prime})\rangle=\frac{k_{B}T}{k}e^{-\lambda|t-t^{\prime}|}\label{eq:autocorr}
\end{equation}
a property guaranteed by Doob's theorem for any Gauss-Markov process
\cite{van1992stochastic}. The Fourier transform of the correlation
function gives the power spectral density 
\begin{equation}
C(\omega)\equiv\langle|x(\omega)|^{2}\rangle=\frac{k_{B}T}{k}\frac{1}{\omega^{2}+\lambda^{2}}\label{eq:spectral-density}
\end{equation}
which is Lorentzian in the angular frequency $\omega$. The corner
frequency $f_{c}=\lambda/2\pi$ is proportional to the mean regression
rate. 

The stationary distribution $P_{1}(x)$ obeys the steady state Fokker-Plank
equation $\mathcal{L}P_{1}=0$ and the solution is, again, a normal
distribution,

\begin{equation}
x\sim\mathcal{N}\left(0,\frac{D}{\lambda}\right)=\mathcal{N}\left(0,\frac{k_{B}T}{k}\right).
\end{equation}
Comparing the forms of $P_{1|1}$ and $P_{1}$ it is clear that former
tends to the latter for $|t-t^{\prime}|\rightarrow\infty$, as it
should. 

The transition probability density yields the Bayesian method for
jointly estimating $\lambda$ and $D$ (and hence $k$), while the
stationary distribution yields the Bayesian method for directly estimating
$k$ alone. We now describe these two methods. 

\section{Bayesian inference\label{sec:Bayesian-inference}}

Consider, now, the time series $X\equiv(x_{1},x_{2},\ldots,x_{N})$
consisting of observations of the sample path $x(t)$ at the discrete
times $t=n\Delta$t with $n=1,\ldots,N.$ Then, using the Markov property
of the Ornstein-Uhlenbeck process, the probability of the sample path
is given by \cite{wang1945theory} 
\begin{equation}
P(X|\lambda,D)=\prod_{n=1}^{N-1}P_{1|1}(x_{n+1}|x_{n},\lambda,D)P_{1}(x_{1}|\lambda,D)
\end{equation}
The probability $P(\lambda,D|X)$ of the parameters, given the sample
path, can now be computed using Bayes theorem, as
\[
P(\lambda,D|X)=\frac{P(X|\lambda,D)P(\lambda,D)}{P(X)}
\]
The denominator $P(X)$ is an unimportant normalization, independent
of the parameters, that we henceforth ignore. Since both $k$ and
$\gamma$ must be positive, for stability and positivity of entropy
production respectively, we use informative priors for $\lambda$
and $D$, $P(\lambda,D)=H(\lambda)H(D)$, where $H$ is the Heaviside
step function. This assigns zero probability weight for negative values
of the parameters and equal probability weight for all positive values.
The logarithm of the posterior probability, after using the explicit
forms of $P_{1|1}$ and $P_{1}$, is
\begin{align}
\ln P(\lambda,D|X) & =\frac{N-1}{2}\ln\frac{\lambda}{2\pi DI_{2}}-\frac{\lambda}{2DI_{2}}\sum_{n=1}^{N-1}\Delta_{n}^{2}\nonumber \\
 & +\frac{1}{2}\ln\frac{\lambda}{2\pi D}-\frac{\lambda}{2D}x_{1}^{2}\label{eq:joint-posterior}
\end{align}
where we have defined the two quantities 
\[
I_{2}\equiv1-e^{-2\lambda\Delta t},\quad\Delta_{n}\equiv x_{n+1}-e^{-\lambda\Delta t}x_{n}.
\]

The maximum a posteriori (MAP) estimate $(\lambda^{\star},D^{\star})$
solves the stationary conditions $\partial\ln P(\lambda,D|X)/\partial\lambda=0$
and $\partial\ln P(\lambda,D|X)/\partial D=0$, while the error bars
of this estimate are obtained from the Hessian matrix of second derivatives
evaluated at the maximum \cite{jeffreys1998theory,jaynes2003probability,sivia2006data}.
The analytical solution of the stationary conditions, derived in the
Appendix, yields the MAP estimate to be\begin{subequations}\label{eq:map-estimate}
\begin{eqnarray}
\lambda^{\star} & = & \frac{1}{\Delta t}\ln\frac{\sum x_{n}^{2}}{\sum x_{n+1}x_{n}}\\
D^{\star} & = & \frac{\lambda^{\star}}{N}\left(\frac{\sum\Delta_{n}^{2}}{I_{2}}+x_{1}^{2}\right)\\
\frac{k^{\star}}{k_{B}T} & = & \frac{\lambda^{\star}}{D^{\star}}
\end{eqnarray}
\end{subequations}where both $I_{2}$ and $\Delta_{n}$ are now evaluated
at $\lambda=\lambda^{\star}$ and the sum runs from $n=1,\ldots,N-1$.
These provide direct estimates of the parameters \emph{without} the
need for fitting, minimization, or Monte Carlo sampling. 

The error bars are obtained from a Taylor expansion of the log posterior
to quadratic order about the MAP value,
\begin{equation}
\ln\frac{P(\lambda,D|X)}{P(\lambda^{\star},D^{\star}|X)}\approx-\left(\Delta\lambda,\Delta D\right)^{T}\mathbf{\boldsymbol{\Sigma}}^{-1}\left(\Delta\lambda,\Delta D\right)\label{eq:quadratic-form}
\end{equation}
where $\Delta\lambda=\lambda-\lambda^{\star}$ and $\Delta D=D-D^{\star}$
and $\boldsymbol{\Sigma}^{-1}$ is the matrix of second derivatives
of the log posterior evaluated at the maximum. The elements $\sigma_{\lambda}^{2}$,
$\sigma_{\lambda D}^{2}$, $\sigma_{D}^{2}$ of the covariance matrix
$\boldsymbol{\Sigma}$ are the Bayesian error bars; they determine
the size and shape of the Bayesian credible region around the maximum
\cite{sivia2006data}. Their unwieldy expressions are provided in
the Appendix and are made use of when computing credible regions around
the MAP estimates. We refer to this Bayesian estimation procedure
as ``Bayes I'' below. 

A second Bayesian procedure for directly estimating the trap stiffness
results when $X$ is interpreted not as a time series but as an exchangeable
sequence, or, in physical terms, as repeated independent observations
of the stationary distribution $P_{1}(x)$ \cite{jaynes2003probability}.
In that case, the posterior probability, assuming an informative prior
that constrains $k$ to positive values, is
\begin{equation}
\ln P(k|X)=\frac{N}{2}\ln\frac{k}{2\pi k_{B}T}-\frac{1}{2}\frac{k}{k_{B}T}\sum_{n=1}^{N}x_{n}^{2}\label{eq:k-posterior}
\end{equation}
The MAP estimate and its error bar follow straightforwardly from the
posterior distribution as
\begin{equation}
\frac{k^{\star}}{k_{B}T}=\frac{N}{\sum_{n=1}^{N}x_{n}^{2}},\quad\sigma_{k}=\frac{1}{\sqrt{N}}\frac{k^{\star}}{k_{B}T}\label{eq:k-map}
\end{equation}
and, not unexpectedly, the standard error decreases as the number
of observations increases. This procedure is independent of $\Delta t$
and is equivalent to the equipartition method when the Heaviside prior
is used for $k$. We refer to this procedure as ``Bayes II'' below.

The posterior probabilities in both methods can be written in terms
of four functions of the data
\begin{eqnarray}
T_{1}(X) & = & \sum_{n=1}^{N-1}x_{n+1}^{2},\quad T_{2}(X)=\sum_{n=1}^{N-1}x_{n+1}x_{n},\nonumber \\
T_{3}(X) & = & \sum_{n=1}^{N-1}x_{n}^{2},\qquad T_{4}(X)=x_{1}^{2},
\end{eqnarray}
which, therefore, are the sufficient statistics of the problem. The
\emph{entire} information in the time series $X$ relevant to estimation
is contained in these four statistics \cite{jaynes2003probability}.
Their use reduces computational expense greatly, as only four numbers,
rather than the entire time series, is needed for evaluating the posterior
distributions.

The posterior distributions obtained above are for flat priors. Other
choice of priors are possible. In particular, since both $D$ and
$k$ are scale parameters a non-informative Jeffreys prior is appropriate
\cite{jeffreys1998theory}. Jeffreys has observed, however, that ``An
accurate statement of the prior probability is not necessary in a
pure problem of estimation when the number of observations is large.''
\cite{jeffreys1998theory}. The number of observations are in the
tens of thousands in time series we study here and the posterior is
dominated by the likelihoood rather than the prior. The prior, then,
has an insignificant contribution to the posterior. 

We note that the error bars obtained in both Bayes I and Bayes II
refer to Bayesian credible intervals, which are relevant to the uncertainty
in the parameter estimates, given the data set $X$. In contrast,
conventional error bars refer to frequentist confidence intervals,
which are relevant to the outcomes of hypothetical repetitions of
measurement. In general, Bayesian credible intervals and frequentist
confidence intervals are not identical and should $not$ be compared
as they provide answers to separate questions \cite{jaynes1976confidence}. 

A comparison of the estimates for the trap stiffness obtained from
these independent procedures provides a check on the validity of the
Ornstein-Uhlenbeck process as a data model. Any significant disagreement
between the estimates from the two methods signals a breakdown of
the applicability of the model and the assumptions implicit in it:
overdamped dynamics, constant friction, harmonicity of the potential,
and thermal equilibrium. This completes our description of the Bayesian
procedures for estimating $\lambda$, $D$, and $k$. 
\begin{figure}[t]
\hfill{}\includegraphics[scale=0.3]{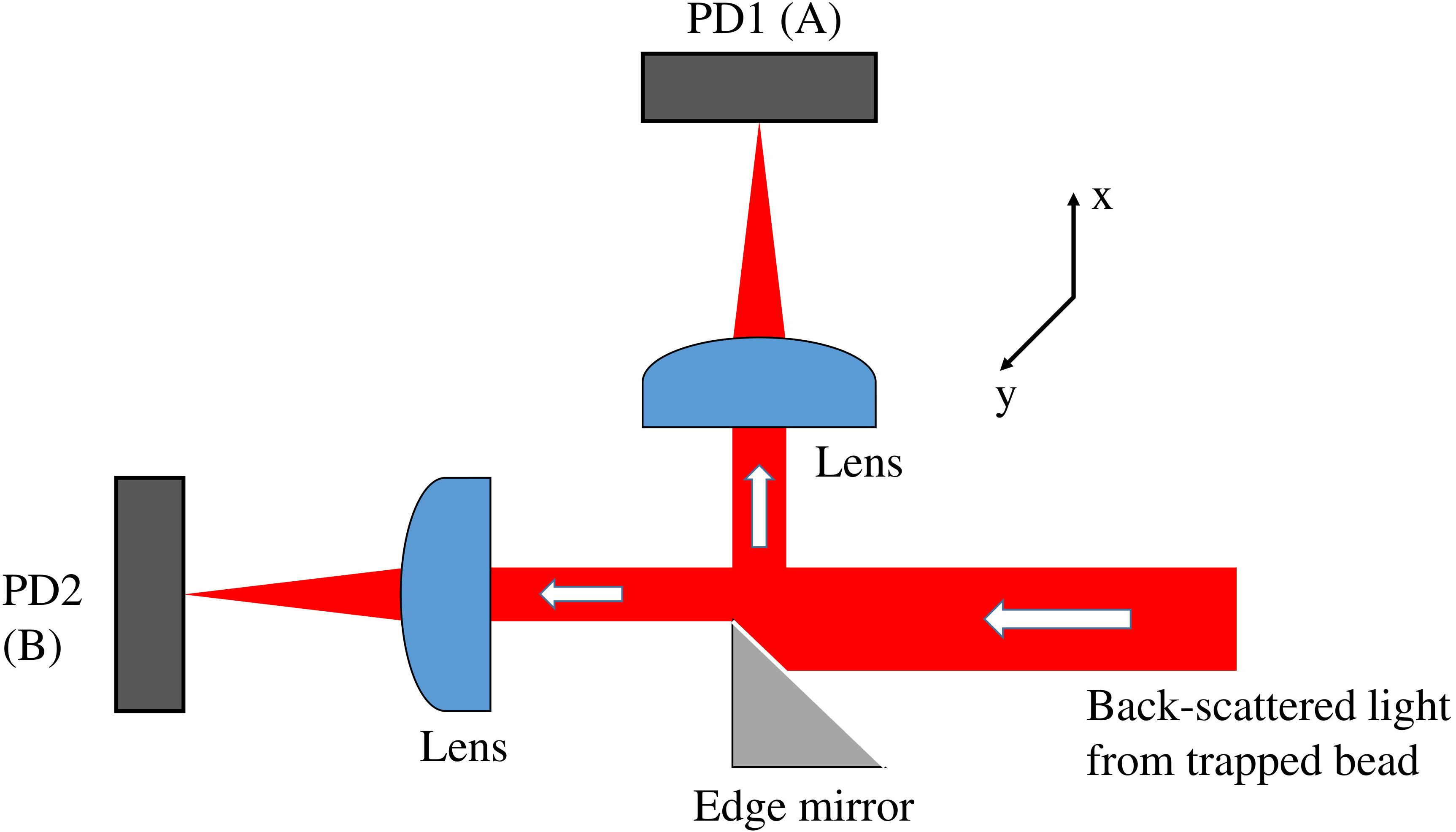}\hfill{}\caption{Schematic of balanced detection scheme to measure Brownian motion
in the $x$ direction from a single trapped polystyrene sphere. Back-scattered
light from the trapped sphere is incident on an edge mirror that divides
it equally between photodiodes PD1 and PD2, having voltage outputs
A and B respectively. The normalized $x$ coordinate of the sphere
at any instant in time is given by $(A-B)/(A+B).$\label{fig:balanced-detection}}
\end{figure}

\section{Experimental setup and data acquisition\label{sec:Experimental-setup-and}}

\begin{figure*}[t]
\includegraphics[scale=0.7]{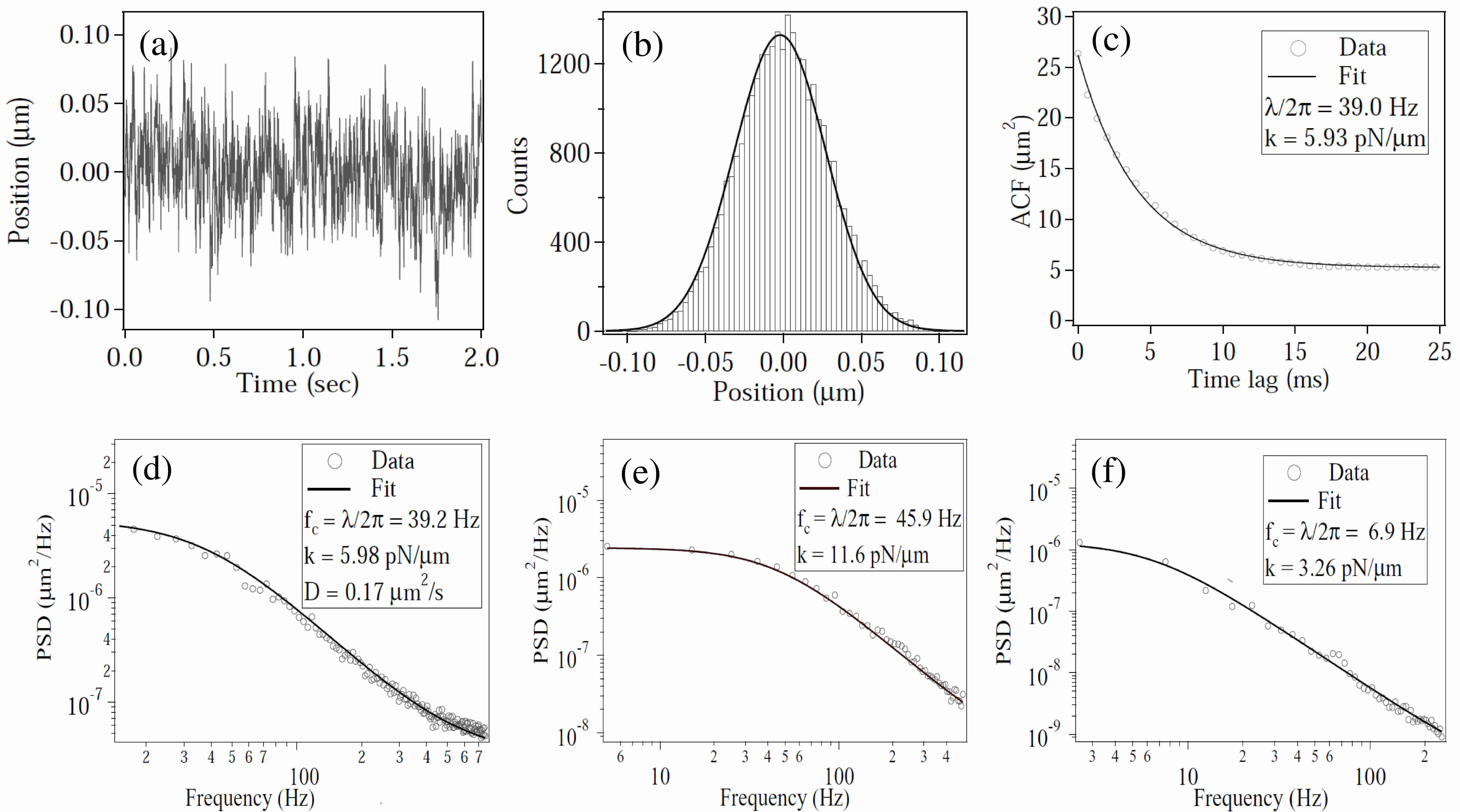}

\caption{Discrete sample path and empirical statistics of an optically trapped
Brownian polystyrene bead of radius $a=3\mu$m. Panel (a) shows discrete
observations of one coordinate of the sample path, (b) the histogram
of the position coordinate, (c) the autocorrelation function and (d)
is the spectral density. Panel (e) and (f) are the spectral density
for 5 and 10 $\mu$m diameter polystyrene spheres. The fits of $\lambda$
from the both the autocorrelation and the spectral density depend,
respectively, on the number of lags and the number of frequencies
used. The guidelines in \cite{berg2004power} and \cite{tassieri2012microrheology}
are followed in obtaining the fits. \label{fig:discrete-sample-path}}
\end{figure*}
We collect position fluctuation data of an optically trapped Brownian
particle using a standard optical tweezers setup that is described
in detail in \cite{rsi12}. Here we provide a brief overview. The
optical tweezers system is constructed around a Zeiss inverted microscope
(\emph{Axiovert.A1}) with a $100x$ $1.4$ numerical aperture (NA)
objective lens tightly focusing laser light at $1064$ nm from a semiconductor
laser (\emph{Lasever}, maximum power $500$mW) into the sample. The
back aperture of the objective is slightly overfilled to maximize
the trapping intensity. The sample consists of a dilute suspension
(volume fraction $\phi=0.01$) of polystyrene spheres of diameter
$3\mu$m in 10\% NaCl-water solution, around $20\mu l$ of which is
pipetted on a standard glass cover slip. The total power available
at the trapping plane is around 15 mW. A single particle is trapped
at an axial distance greater than several times its radius to avoid
any wall-effects in the effective drag force due to the water, and
it's motion is observed by back-focal plane interferometry using the
back-scattered intensity of a detection laser at $671$ nm that co-propagates
with the trapping laser. The detection laser power is maintained at
much lower levels than that required to trap a particle. The back-scattered
signal from the trapped particle is measured using a balanced detection
scheme, schematically illustrated in Fig. \ref{fig:balanced-detection}.
The back-scattered light beam is incident on an edge mirror which
divides it equally into two halves that are focused using two lenses
of equal focal length on photodiodes PD1 and PD2 (\emph{Thorlabs}
PDA100A-EC Si-photodiodes of bandwidth 2.4 MHz). The voltage outputs
$A$ and $B$, of PD1 and PD2 respectively, are then combined as $(A-B)/(A+B)$
to give the normalized value of the $x$ coordinate of motion at any
instant of time. The advantage of such balanced detection is that
the intensity fluctuations of the laser are present in both beams
simultaneously and are thus canceled out when the difference is taken.
Note that the direction of the edge mirror decides whether the $x$
or $y$ coordinate of motion is being measured. The mirror is rotated
by $90$ degrees to select between the coordinates. The fast response
of the photodiodes, with a rise time of 50ns at highest gain, ensures
that spurious correlations are kept to a minimum and the data filtering
necessary with slower commercial quadrant photodetectors is avoided
entirely. The data from the photodiodes is logged
into a computer using a National Instruments DAQ system and Labview
at sampling rates between 2-5 kHz. For calibrating the motion, \emph{i.e.}
converting the voltage into physical distance which is necessary for
measuring the diffusion constant, we employ an acousto-optic modulator
that is placed in the back-focal plane of the microscope objective
and scan the trapped bead by distances which are determined from the
pixel calibration of images taken by the camera attached to the microscope
\cite{rsi12}. The balanced detection output is simultaneously measured
to yield the voltage-distance calibration of the detection system.
The detection signal amplitude for Brownian motion
data for 3 $\mu$m diameter spheres in water is around 1.5 V/$\mu$m
and the noise floor is around 5 mV, which implies that we have a sensitivity
of around 7 nm (considering signal/noise=2) for this case. However,
since scattering from spheres depends on diameter (generally increasing
with diameter) as well as the refractive index of the ambient medium,
this value changes when we change spheres or the medium. Typically,
the particle localization is within 2-7 nm in our experiments. For
the viscosity measurement, we add glycerol to water in fixed proportions
to create 5 samples of different viscosity. The viscosity of each
sample is then measured by a commercial rheometer (Brookfield DB3TLVCJ0)
to match with the experimental results. The voltage-distance
calibration is performed every time we change the particle or the
ambient medium.
\begin{figure}[h]
\centering\includegraphics[width=0.46\textwidth]{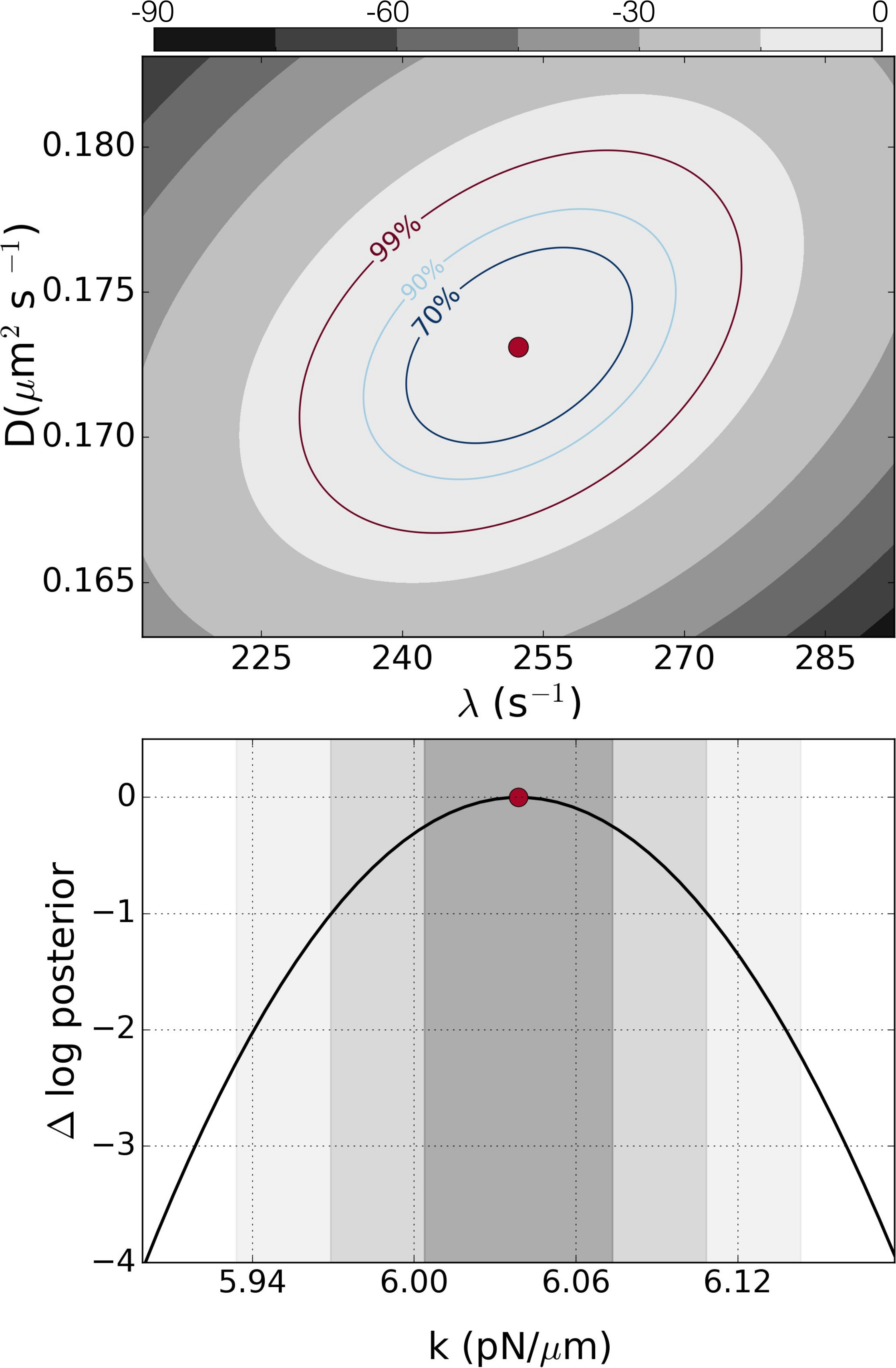}

\caption{Bayesian posterior probability densities. The top panel shows filled
contours of Eq. (\ref{eq:joint-posterior}). The MAP estimate, Eq.
(\ref{eq:map-estimate}), is marked by the filled dot and contours
enclosing 70\%, 90\% and 99\% of the probability are labeled. The
bottom panel shows Eq. (\ref{eq:k-posterior}). The MAP estimate,
Eq. (\ref{eq:k-map}), is marked by the filled dot and intervals enclosing
70\%, 90\% and 99\% of the probability are shaded. The two estimates
for $k$ agree to three decimal places.\label{fig:bayes-I-and-II}}
\end{figure}
\begin{figure}[h]
\centering\includegraphics[width=0.46\textwidth]{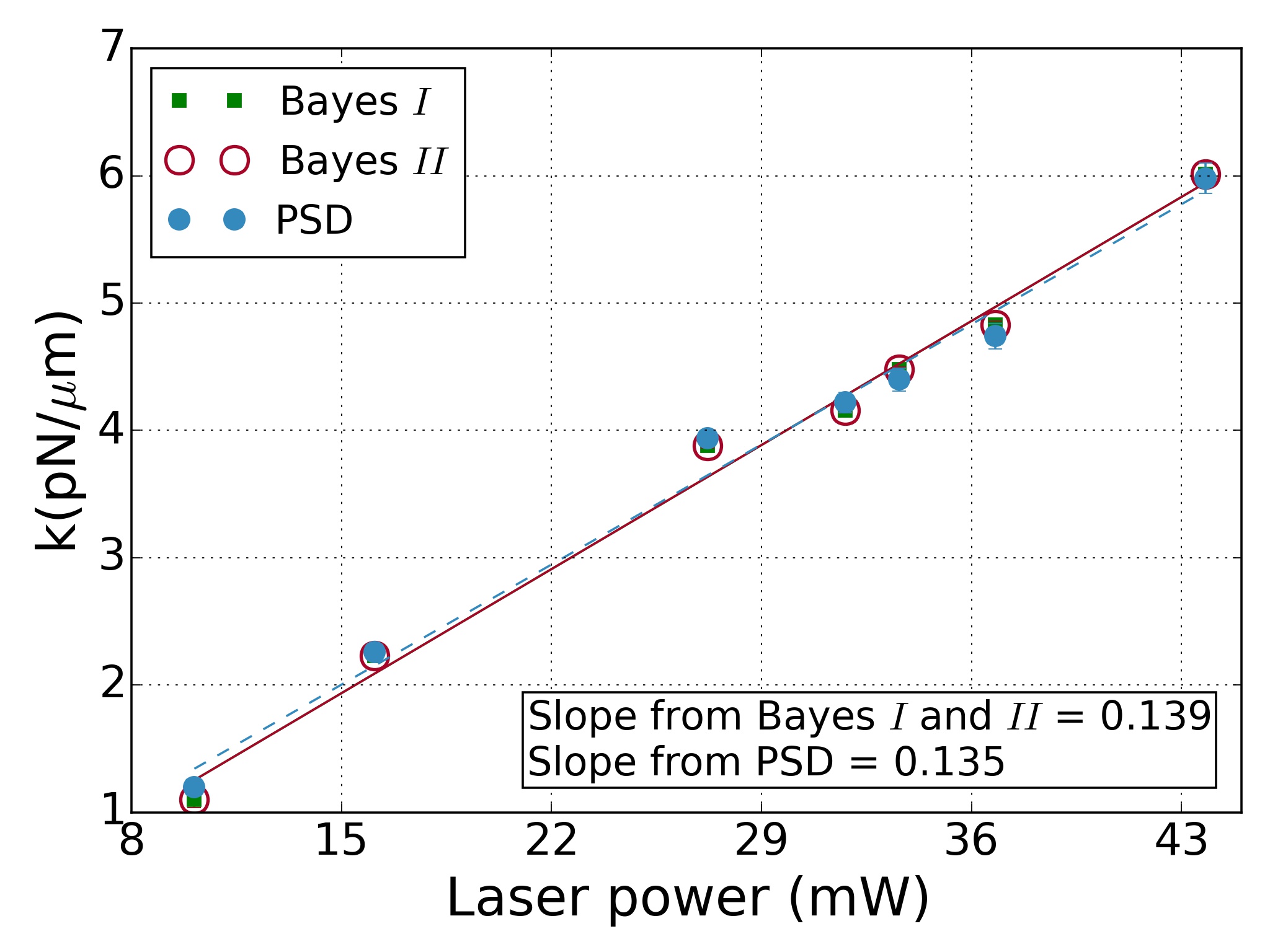}

\caption{Variation of trap stiffness $k$ with laser power estimated by the
two Bayesian methods (Bayes I and Bayes II) and by the standard fit
to the power spectral density (PSD). The error bars are also shown.
Solid line is the best fit to Bayes I and II while the dotted line
is a best fit to PSD.\label{fig:Variation-of-trap}}
\end{figure}

We note that the measured data is a result of a transformation by
the detection apparatus of the physical sample paths. The Bayesian
modeling of the detection apparatus and the transformations it induces
on the physical sample paths is not pursued here. Therefore, we have
a problem of pure estimation and there is no attempt to compare between
alternative models of the data generation process. 
\begin{table}
\begin{tabular}{>{\centering}p{2cm}>{\centering}p{2cm}>{\centering}p{2cm}>{\centering}p{2cm}}
\toprule 
\multirow{2}{2cm}{Laser power (mW) } & \multicolumn{3}{c}{$k$ ($pN\mu m^{-1}$)}\tabularnewline
\cmidrule{2-4} 
 & Bayes I & Bayes II & PSD\tabularnewline
\midrule
10.1 & 1.10(6) & 1.10(6) & 1.20(2)\tabularnewline
\midrule
16.1 & 2.23(1) & 2.23(5) & 2.26(5) \tabularnewline
\midrule
27.2 & 3.88(2) & 3.88(5) &  3.94(6)\tabularnewline
\midrule
31.8 & 4.16(2) & 4.16(2) & 4.22(8)\tabularnewline
\midrule
33.6 & 4.48(2) & 4.48(2) & 4.40(9)\tabularnewline
\midrule
36.8 & 4.83(3) & 4.83(2) & 4.74(10)\tabularnewline
\midrule
43.8 & 6.01(3) & 6.01(3) & 5.98(12)\tabularnewline
\bottomrule
\end{tabular}\caption{Variation of trap stiffness $k$ with laser power estimated by the
two Bayesian methods of this work (Bayes I and Bayes II) and by the
standard fit to the spectral density (PSD) \cite{berg2004power}.
The variance of mean is indicated in parentheses. The Bayesian standard
error is less than $1\%$ of the mean for each data set. \label{tab:k-versus-power}}
\end{table}
\begin{table*}
\begin{tabular}{>{\centering}p{2cm}>{\centering}p{2cm}>{\centering}p{2cm}>{\centering}p{2cm}>{\centering}p{2cm}>{\centering}p{2cm}>{\centering}p{2cm}}
\toprule 
\multirow{3}{2cm}{Particle diameter ($\mu m$)} & \multicolumn{3}{c}{$k$ ($pN\mu m^{-1}$)} & \multicolumn{3}{c}{$D^{\star}$ ($10^{-13}ms^{-1}$)}\tabularnewline
 &  &  &  &  &  & \tabularnewline
\cmidrule{2-7} 
 & Bayes I & Bayes II & PSD & Bayes I & Bayes II & PSD\tabularnewline
\midrule
3 & 6.01(3) & 6.01(3) & 5.98(12) & 1.73(6) & 1.74(6) & 1.70(8)\tabularnewline
\midrule
5 & 11.5(1) & 11.5(1) & 11.6(3)  & 1.015(10) & 1.014(10) & 1.03(5) \tabularnewline
\midrule
10 & 3.03(2) & 3.03(5) &  3.26(6) & 0.505(5) & 0.504(5) &  0.50(1)\tabularnewline
\bottomrule
\end{tabular}\caption{\label{tab:k-different-dia}Trap stiffness $k$ and diffusion constant
$D^{\star}$ measured for 3, 5, and 10 $\mu$m diameter polystyrene
spheres trapped at the same laser power. The variance of mean is indicated
in parentheses. }
\end{table*}
\section{Results and discussions\label{sec:Results}}
We now present our results. In Figure. (\ref{fig:discrete-sample-path})
we show a typical sample path of one component of motion in the plane
of the trap, together with its histogram, autocorrelation function
and spectral density. The histogram shows that the distribution of
positions is stationary and very well-approximated by a Gaussian.
The variance $\langle x^{2}\rangle$ is used in the conventional ``equipartition''
method to estimate the spring constant $k$, while the fitting of
the autocorrelation to the exponential in Eq.(\ref{eq:autocorr})
or of the spectral density to the Lorentzian in Eq. (\ref{eq:spectral-density})
is used to estimate the spring constant when the friction constant
is given. For estimation of the stiffness from the PSD, we employ
the procedures suggested in \cite{berg2004power}, including ``blocking''
the data with a bin size of 100 points, and setting the frequency
range for fitting in order to avoid systematic errors due to reliability
issues at both low and high frequencies \cite{berg2004power}. However,
there exist issues in estimating stiffness from both the equipartition
method - where the presence of any additive noise leads to an increase
in the variance that leads to over-estimation of the trap stiffness,
and the PSD - where the standard systematics related to fitting can
be minimized at best but not removed. 

The results of Bayesian inference are shown in Figure. (\ref{fig:bayes-I-and-II}).
In the top panel we show filled contours of the posterior distribution
in the $\lambda-D$ plane, together with contours of equal probability,
for the ``Bayes I'' method. There is a single maximum at $(\lambda^{\star},D^{\star})$
whose numerically computed value is in excellent agreement with the
analytical MAP estimates of Eq.(\ref{eq:map-estimate}). In the bottom
panel we show the Bayesian posterior distribution for the stiffness
for the ``Bayes II'' method. There is remarkably good agreement
between the two Bayesian estimates and the fit to the power spectral
density as shown in Fig. (\ref{fig:Variation-of-trap}) and Table.
(\ref{tab:k-versus-power}). This consistency between three conceptually
and procedurally independent methods is evidence for the appropriateness
of the data model. The agreement with the spectral density method,
shown in the third column of the table, is within $2-3\%$ in all
cases, other than the first case where the inherent low stiffness
of the trap due to low trapping power led to larger systematics due
to the increased influence of the ambient low frequency noise, as
we shall discuss later. The typical length of our time series is
$N\sim30000$ and this gives a Bayesian error bar that is less than
$\frac{1}{2}$\% of the mean. These are well below the systematic
errors and the approximately $1.5-2\%$ variability of the estimates
obtained from the fitting procedure. The $1\sigma$ standard errors
in the mean are indicated in parenthesis next to each $k$ value in
Table. (\ref{tab:k-versus-power}). Note that for inference of the
absolute values of $k$ and $D^{\star}$, we have used a temperature
of 300K which is the same as the lab environment temperature -
this assumption being based on studies in literature \cite{schmidt03heating}where
the effects of laser heating in water has shown to be well below 1K
at the power levels we employ in the trap. However, the fact that
the results from Bayes I and Bayes II are extremely close to each
other demonstrates that our estimate of temperature is trustworthy.
Next, in order to determine the robustness of our
experimental techniques, we perform experiments on two other sizes
of polystyrene spheres - of diameter 5 and 10 $\mu$m, respectively.
Representative values of measured stiffness and diffusion constant
at a laser power of 43.8 mW are shown in Table \ref{tab:k-different-dia}.
It is clear that the remarkable consistency in the values of $k$
and $D^{\star}$ given by the Bayes I and II methods are preserved,
as is the close agreement with the values obtained from the PSD analysis.
Note that the stiffness values are not really related to bead size
as is well known in literature - the dependence of $k$ on bead diameter
being rather non-linear \cite{simmonsbeadstiff96}. Our next set of measurements are directed towards
determining the extent of systematic errors in our measurements. Systematic
errors in our experimental apparatus may arise due to various issues
including slow drifts of the laser power, beam pointing of the laser,
drifts in ambient temperature, coupling with ambient low frequency
(typically acoustic) noise sources, and possibly other unidentified
reasons. The coupling with ambient low frequency noise sources would
manifest themselves at low trapping stiffness, where sudden perturbations
could affect the Brownian motion of the bead since the restoring force
is less. On the other hand, the affect of slow drifts of experimental
parameters would be observed at time series data of longer length.
Thus, we attempt to understand the effect of systematics in our experiments
using two different approaches: a) by comparing the mean and standard
deviation of measured trapping parameters over 3 sets of independent
time series data collected at low and high trapping laser powers for
the same particle (diameter 3 $\mu$m), and b) by comparing the trapping
parameters measured on time series data of different length obtained
at the same laser power for the same particle. The results are shown
in Table \ref{tab:systematic-error}(a) and (b). In Table \ref{tab:systematic-error}(a),
we observe that at a low laser power of 18.5 mW, we have a standard
deviation of around 6.5\% in $k$
and 4.5\% in $D^{\star}$, while at the higher power of 43.8 mW, the
standard deviation is only around 1.5\% for both$k$
and $D^{\star}$. This demonstrates that the effect
of ambient noise does increase at lower trapping powers. In Table
\ref{tab:systematic-error}(b), we show the results of measurements
of $k$ and $D^\star$for time series
data of different lengths for the same particle trapped at a laser
power of 48 mW (we choose a high laser power since the coupling with
the ambient noise is lesser in that case). It is clear that for time
series of lengths 5, 10, and 20s, the variation in the mean of both
$k$ and $D^{\star}$is only around
1.5\%, but there is a large change in the mean values between 7-24\%
for time series data of 40 and 60s. In addition, we check that the
estimates of $k$ and $D^{\star}$over
different non-overlapping segments of up to 20s length in a \emph{single
}data set are within the Bayesian error bars for
all the time series; however, for 40 and 60s, the mean values of different
non-overlapping sets of 20s data differ from each other significantly.
Thus, it is clear that systematics due to slow drifts in different
experimental apparatus occur at time scales longer than 20s. Note
that we perform our experiments on an optical table with active vibration
isolation, so that there is no coupling with ambient vibrations, whereas
vibrations from the table itself are at much higher frequencies than
our region of interest due to the large table mass and are also damped
out very fast by the presence of active dampers (we also take care
not to place vibrating objects such as power supplies, etc on the
table).
\begin{table*}
\begin{tabular}{>{\centering}m{2cm}>{\centering}p{2cm}>{\centering}p{2.5cm}>{\centering}p{1cm}}
\cmidrule{1-3} 
Laser power
(mW) & $k$ ($pN\mu m^{-1}$) & $D^{\star}$ ($10^{-13}ms^{-1}$) & \tabularnewline
\cmidrule{1-3} 
18.5 & 2.44(16) & 1.74(8) & \tabularnewline
\cmidrule{1-3} 
43.8 & 5.98(9) & 1.75(3)  & \tabularnewline
\cmidrule{1-3} 
 & (a) &  & \tabularnewline
\end{tabular}%
\begin{tabular}{>{\centering}m{2cm}>{\centering}p{2cm}>{\centering}p{2.5cm}}
\toprule 
Time series length (s) & $k$ ($pN\mu m^{-1}$) & $D^{\star}$ ($10^{-13}ms^{-1}$)\tabularnewline
\midrule 
5 & 6.77(6) & 1.78(4)\tabularnewline
\midrule 
10 & 6.58(5) & 1.74(3) \tabularnewline
\midrule 
20 & 6.72(4) & 1.73(2)\tabularnewline
\midrule 
40 & 6.29(2) & 2.01(1)\tabularnewline
\midrule 
60 & 5.10(1) & 1.93(1)\tabularnewline
\midrule
 & (b) & \tabularnewline
\end{tabular}\caption{\label{tab:systematic-error}Study of systematic error in the experimental
apparatus by analysis of particle trajectory for (a) same particle
trapped at different laser powers, and (b) time series of different
lengths for the same particle trapped at the same laser power. The
variance of mean is indicated in parentheses. }
\end{table*}
\begin{table}
\begin{tabular}{>{\centering}m{2cm}>{\centering}p{2cm}>{\centering}p{2cm}>{\centering}p{2cm}}
\toprule 
$\eta$ & $D$ & $D^{\star}$ & $\eta^{\star}$\tabularnewline
\midrule 
0.00085 & 1.72 & 1.73(6) & 0.00084(3)\tabularnewline
\midrule 
0.00089 & 1.65 & 1.72(6)  & 0.00085(3)\tabularnewline
\midrule 
0.00137 & 1.07 & 1.05(3) & 0.00139(4)\tabularnewline
\midrule 
0.00197 & 0.743 & 0.732(11) & 0.00200(3)\tabularnewline
\midrule 
0.00243 & 0.603 & 0.586(12) & 0.00250(5)\tabularnewline
\midrule 
0.00487 & 0.301 & 0.276(14) & 0.00530(24)\tabularnewline
\bottomrule
\end{tabular}\caption{Bayesian viscometry in an optical trap. The first column is the viscosity
of the solvent as measured in a rheometer and the second column is
the diffusion coefficient as given by the Stokes-Einstein relation
for that value of the viscosity. The third column is the Bayesian
MAP estimate for the diffusion coefficient and the fourth column is
the value of the viscosity, as given by the Stokes-Einstein relation
for the corresponding value of the diffusion coefficient. There is
a good match between the first and fourth columns. Note that the first
row is for water while the rest are for water + glycerol samples with
increasing glycerol concentration. The variance of mean is indicated
in parentheses. \label{tab:viscometry}}
\end{table}

To compare the Bayesian estimate for the diffusion coefficient we
repeat the experiment for different solvent viscosities keeping both
the laser power ( corresponding to $k\sim$ 6 $pN$) and the particle
radius ($a=3\mu m$) fixed. The Stokes-Einstein relation then provides
an estimate of the diffusion coefficient. We compare this estimate
with the MAP estimate $D^{\star}$ in Table. (\ref{tab:viscometry})
to find agreement to within $10\%$ in all cases. The Stokes-Einstein
relation can be used ``in reverse'' to obtain a MAP estimate of
the viscosity, $\eta^{\star}$, which agrees very well with the known
viscosity of the mixture. The experiments were performed for five
sets of data for each viscosity sample and the mean value of $D^{\star}$
has been reported with the corresponding 1$\sigma$ error in parenthesis.
The error bars, which are higher than that for the $k$ measurement,
chiefly reflect the systematic errors in our experimental apparatus
that have been described previously and occur at
time scales longer than $20s$, which is the duration over which a
single data set is collected. As mentioned earlier,
we have checked that the estimates of $D^{\star}$over different
non-overlapping segments of a \emph{single }data set are within the
Bayesian error bars, which again confirms that the
errors we observe are due to systematic shifts in the operating conditions
of the experiment. The agreement in the viscosity values for that
measured in the rheometer and by the Bayesian estimate of $D^{\star}$is
within 5\% for all cases with the exception of the last, where the
enhanced friction caused a shift in $\lambda$ towards lower values,
once again increasing the effects of systematics due to ambient low
frequency noise sources. Also, since this value corresponded to the
highest concentration of glycerol in the water + glycerol mixture,
the effects of laser heating could have been more significant \cite{schmidt03heating},
leading to a slight increase of temperature at the trap focal volume
which we have not considered in the Bayesian estimate. This \textquotedblleft fluctuation\textquotedblright{}
method of estimating the viscosity does not require the application
of external fields and is so guaranteed to yield the linear response
of the system while the Bayesian analysis extracts, optimally, all
information relevant to this estimation problem. The viscosity of
nanoliter samples can be estimated by this method, making it an attractive
alternative to \textquotedblleft response\textquotedblright{} methods
that impose an external shear flow.
\section{Conclusion\label{sec:Conclusion}}
In this work, we have presented an exact Bayesian method for jointly
estimating the mean regression rate and the diffusion coefficient
of an optically trapped Brownian particle. The trap stiffness in temperature
units is obtained as a ratio of the mean regression rate and the diffusion
coefficient. We have also rephrased the standard ``equipartition''
method of directly estimating the trap stiffness as a problem in Bayesian
inference. We have assumed that the Ornstein-Uhlenbeck process is
the data generating model. More general models, which include the
position dependence of the particle friction (as would be the case
in the proximity to walls) or the non-Markovian character of the trajectories
(as would be the case when momentum diffusion is not slow compared
to the time scales of interest) can, with additional effort, be incorporated
in the Bayesian framework. Exact analytical solutions will no longer
be available and one has to resort to approximations of the likelihood,
such as short-time expansions of the Fokker-Planck propagator or numerical
solutions of the equivalent stochastic differential equations. These
introduce discretization errors which must be carefully evaluated.
In contrast, the method presented here is exact and can serve as an
useful ``null hypothesis'' when comparing between different models
for the data. In future work, we shall present Bayesian methods for
more complex models and provide a fully Bayesian procedure, embodying
Ockham's razor \cite{jaynes2003probability}, for the problem of model
selection. 

Bayesian analysis is generally applicable in studying the dynamics
(Brownian or otherwise) of a vast range of mesoscopic particles in
diverse trapping environments. While we have focussed
on spherical particles here, the method is not restricted thus, and
can be applied to non-spherical particles. The rotational and translation
motions are typically coupled in such cases, making the analysis more
difficult in detail but no different in principle. The capability
of modeling the apparatus itself could also be extremely helpful in
understanding and improving the experimental techniques employed in
optical trapping with the possibility of studying different systematic
effects that may influence the trajectory of trapped particles, thus
enhancing the capabilities and sphere of influence of optical tweezers.

Bayesian methods for data analysis are not widespread in soft matter,
despite of their advantages and demonstrated success in other areas
of physics. To quote a popular textbook \cite{gregory2005bayesian}:
``Increasingly, researchers in many branches of science are coming
into contact with Bayesian statistics or Bayesian probability theory.
By encompassing both inductive and deductive logic, Bayesian analysis
can improve model parameter estimates by many orders of magnitude.
It provides a simple and unified approach to all data analysis problems,
allowing the experimenter to assign probabilities to competing hypotheses
of interest, on the basis of the current state of knowledge.'' In
future, we hope to see many more applications of this ``elegant and
powerful approach to scientific inference''\cite{gregory2005bayesian}
to problems in soft matter.
\begin{acknowledgments}
RA expresses his gratitude, long overdue, to Professor M. E. Cates
for introducing him to Bayesianism.
\end{acknowledgments}
\subsection{Appendix }
The first partial derivatives of the logarithm of the posterior probability
with respect to $\lambda$ and $D$ are
\begin{alignat*}{1}
\frac{\partial\ln P}{\partial\lambda} & =\frac{N-1}{2}\left(\frac{1}{\lambda}-\frac{I_{2}^{\prime}}{I_{2}}\right)-\frac{\sum\Delta_{n}^{2}}{2DI_{2}}\\
 & -\frac{\lambda}{2D}\frac{\partial}{\partial\lambda}\left(\frac{\sum\Delta_{n}^{2}}{I_{2}}\right)+\frac{1}{2\lambda}-\frac{x_{1}^{2}}{2D},\\
\frac{\partial\ln P}{\partial D} & =-\frac{N-1}{2D}+\frac{\lambda\sum\Delta_{n}^{2}}{2D^{2}I_{2}}-\frac{1}{2D}+\frac{\lambda x_{1}^{2}}{2D^{2}},
\end{alignat*}
where $I_{2}^{\prime}=2\Delta t\thinspace e^{-2\lambda\Delta t}$.
Setting the second of these equations to zero, $D$ is solved in term
of $\lambda$ and this solution is used in the first equation, together
with the large-sample asymptotics
\begin{alignat*}{1}
\frac{\lambda}{N}\left(\frac{\sum\Delta_{n}^{2}}{I_{2}}+x_{1}^{2}\right) & \thickapprox\frac{\lambda}{\left(N-1\right)}\frac{\sum\Delta_{n}^{2}}{I_{2}},
\end{alignat*}
to cancel all $D$-dependent terms. Setting the resulting equation
to zero and solving for $\lambda$ then yields the MAP estimates in
Eq.(\ref{eq:map-estimate}). 

The second partial derivatives, appearing in Eq. (\ref{eq:quadratic-form}),
are 
\begin{align*}
\Sigma_{11}^{-1}=\frac{\partial^{2}\ln P}{\partial\lambda^{2}}= & \frac{N-1}{2}\Big(-\frac{1}{\lambda^{2}}+\frac{I_{2}^{\prime}I_{2}^{\prime}}{I_{2}^{2}}-\frac{I_{2}^{\prime\prime}}{I_{2}}\Big)-\frac{1}{2\lambda^{2}}\\
- & \frac{1}{D}\frac{\partial}{\partial\lambda}\left(\frac{\sum\Delta_{n}^{2}}{I_{2}}\right)-\frac{\lambda}{2D}\frac{\partial^{2}}{\partial\lambda^{2}}\left(\frac{\sum\Delta_{n}^{2}}{I_{2}}\right),\\
\Sigma_{12}^{-1}=\frac{\partial^{2}\ln P}{\partial D\partial\lambda}= & \frac{\sum\Delta_{n}^{2}}{2D^{2}I_{2}}+\frac{\lambda}{2D^{2}}\frac{\partial}{\partial\lambda}\left(\frac{\sum\Delta_{n}^{2}}{I_{2}}\right)+\frac{x_{1}^{2}}{2D^{2}},\\
\Sigma_{22}^{-1}=\frac{\partial^{2}\ln P}{\partial D^{2}}= & \frac{N-1}{2D^{2}}-\frac{\lambda}{D^{3}}\Big(\frac{\sum\Delta_{n}^{2}}{I_{2}}-x_{1}^{2}\Big)+\frac{1}{2D^{2}},
\end{align*}
where $I_{2}^{\prime\prime}=-4\Delta t^{2}\thinspace e^{-2\lambda\Delta t}$.
All the derivatives are evaluated at the maximum given in Eq. (\ref{eq:map-estimate}).
These are assembled into the Hessian matrix $\boldsymbol{\Sigma}^{-1}$and
the matrix is inverted to give the covariance matrix $\boldsymbol{\Sigma}$
in Eq. (\ref{eq:quadratic-form}), whose matrix elements are $\sigma_{\lambda}^{2}$,
$\sigma_{\lambda D}^{2}$, $\sigma_{D}^{2}$ 
\begin{eqnarray*}
\sigma_{\lambda}^{2}= &  & -\frac{1}{\det\boldsymbol{\Sigma}^{-1}}\Sigma_{22}^{-1},\qquad\sigma_{\lambda D}^{2}=\frac{1}{\det\boldsymbol{\Sigma}^{-1}}\Sigma_{12}^{-1},\\
\sigma_{D}^{2}= &  & -\frac{1}{\det\boldsymbol{\Sigma}^{-1}}\Sigma_{11}^{-1},
\end{eqnarray*}
where $\det\boldsymbol{\Sigma}^{-1}=\Sigma_{11}^{-1}\Sigma_{22}^{-1}-\Sigma_{12}^{-1}\Sigma_{21}^{-1}$
and $\Sigma_{21}^{-1}=\Sigma_{12}^{-1}$.


\begin{thebibliography}{10}
\expandafter\ifx\csname url\endcsname\relax
  \def\url#1{\texttt{#1}}\fi
\expandafter\ifx\csname urlprefix\endcsname\relax\def\urlprefix{URL }\fi
\expandafter\ifx\csname doiprefix\endcsname\relax\def\doiprefix{DOI }\fi
\providecommand{\bibinfo}[2]{#2}
\providecommand{\eprint}[2][]{\url{#2}}

\bibitem{chandrasekhar1943stochastic}
\bibinfo{author}{Chandrasekhar, S.}
\newblock \bibinfo{title}{Stochastic problems in physics and astronomy}.
\newblock \emph{\bibinfo{journal}{Reviews of modern physics}}
  \textbf{\bibinfo{volume}{15}}, \bibinfo{pages}{1} (\bibinfo{year}{1943}).

\bibitem{van1992stochastic}
\bibinfo{author}{Van~Kampen, N.~G.}
\newblock \emph{\bibinfo{title}{Stochastic processes in physics and
  chemistry}}, vol.~\bibinfo{volume}{1} (\bibinfo{publisher}{Elsevier},
  \bibinfo{year}{1992}).

\bibitem{jeffreys1998theory}
\bibinfo{author}{Jeffreys, H.}
\newblock \emph{\bibinfo{title}{The theory of probability}}
  (\bibinfo{publisher}{OUP Oxford}, \bibinfo{year}{1998}).

\bibitem{zellner1988optimal}
\bibinfo{author}{Zellner, A.}
\newblock \bibinfo{title}{Optimal information processing and bayes's theorem}.
\newblock \emph{\bibinfo{journal}{The American Statistician}}
  \textbf{\bibinfo{volume}{42}}, \bibinfo{pages}{278--280}
  (\bibinfo{year}{1988}).

\bibitem{gardiner1985handbook}
\bibinfo{author}{Gardiner, C.~W.} \emph{et~al.}
\newblock \emph{\bibinfo{title}{Handbook of stochastic methods}},
  vol.~\bibinfo{volume}{3} (\bibinfo{publisher}{Springer Berlin},
  \bibinfo{year}{1985}).

\bibitem{kubo1966fluctuation}
\bibinfo{author}{Kubo, R.}
\newblock \bibinfo{title}{The fluctuation-dissipation theorem}.
\newblock \emph{\bibinfo{journal}{Reports on progress in physics}}
  \textbf{\bibinfo{volume}{29}}, \bibinfo{pages}{255} (\bibinfo{year}{1966}).

\bibitem{berg2004power}
\bibinfo{author}{Berg-S{\o}rensen, K.} \& \bibinfo{author}{Flyvbjerg, H.}
\newblock \bibinfo{title}{Power spectrum analysis for optical tweezers}.
\newblock \emph{\bibinfo{journal}{Review of Scientific Instruments}}
  \textbf{\bibinfo{volume}{75}}, \bibinfo{pages}{594--612}
  (\bibinfo{year}{2004}).

\bibitem{richly2013calibrating}
\bibinfo{author}{Richly, M.~U.} \emph{et~al.}
\newblock \bibinfo{title}{Calibrating optical tweezers with bayesian
  inference}.
\newblock \emph{\bibinfo{journal}{Optics express}}
  \textbf{\bibinfo{volume}{21}}, \bibinfo{pages}{31578--31590}
  (\bibinfo{year}{2013}).

\bibitem{jaynes1976confidence}
\bibinfo{author}{Jaynes, E.~T.} \& \bibinfo{author}{Kempthorne, O.}
\newblock \bibinfo{title}{Confidence intervals vs bayesian intervals}.
\newblock In \emph{\bibinfo{booktitle}{Foundations of probability theory,
  statistical inference, and statistical theories of science}},
  \bibinfo{pages}{175--257} (\bibinfo{publisher}{Springer},
  \bibinfo{year}{1976}).

\bibitem{pybisp}
\bibinfo{author}{Singh, R.} \& \bibinfo{author}{Adhikari, R.}
\newblock \bibinfo{title}{Pybisp}.
\newblock \urlprefix\url{https://github.com/ronojoy/pybisp}.

\bibitem{wang1945theory}
\bibinfo{author}{Wang, M.~C.} \& \bibinfo{author}{Uhlenbeck, G.~E.}
\newblock \bibinfo{title}{On the theory of the brownian motion ii}.
\newblock \emph{\bibinfo{journal}{Reviews of Modern Physics}}
  \textbf{\bibinfo{volume}{17}}, \bibinfo{pages}{323} (\bibinfo{year}{1945}).

\bibitem{jaynes2003probability}
\bibinfo{author}{Jaynes, E.~T.}
\newblock \emph{\bibinfo{title}{Probability theory: The logic of science}}
  (\bibinfo{publisher}{Cambridge university press}, \bibinfo{year}{2003}).

\bibitem{sivia2006data}
\bibinfo{author}{Sivia, D.} \& \bibinfo{author}{Skilling, J.}
\newblock \emph{\bibinfo{title}{Data analysis: a Bayesian tutorial}}
  (\bibinfo{publisher}{OUP Oxford}, \bibinfo{year}{2006}).

\bibitem{rsi12}
\bibinfo{author}{Pal, S.~B.}, \bibinfo{author}{Haldar, A.},
  \bibinfo{author}{Roy, B.} \& \bibinfo{author}{Banerjee, A.}
\newblock \bibinfo{title}{Measurement of probe displacement to the thermal
  resolution limit in photonic force microscopy using a miniature quadrant
  photodetector}.
\newblock \emph{\bibinfo{journal}{Review of Scientific Instruments}}
  \textbf{\bibinfo{volume}{83}}, \bibinfo{pages}{023108}
  (\bibinfo{year}{2012}).

\bibitem{tassieri2012microrheology}
\bibinfo{author}{Tassieri, M.}, \bibinfo{author}{Evans, R.},
  \bibinfo{author}{Warren, R.~L.}, \bibinfo{author}{Bailey, N.~J.} \&
  \bibinfo{author}{Cooper, J.~M.}
\newblock \bibinfo{title}{Microrheology with optical tweezers: data analysis}.
\newblock \emph{\bibinfo{journal}{New Journal of Physics}}
  \textbf{\bibinfo{volume}{14}}, \bibinfo{pages}{115032}
  (\bibinfo{year}{2012}).

\bibitem{schmidt03heating}
\bibinfo{author}{Peterman, E. J.~G.}, \bibinfo{author}{Gittes, F.} \&
  \bibinfo{author}{Schmidt, C.~F.}
\newblock \bibinfo{title}{On the theory of the brownian motion ii}.
\newblock \emph{\bibinfo{journal}{Biophysical Journal}}
  \textbf{\bibinfo{volume}{84}}, \bibinfo{pages}{1308} (\bibinfo{year}{2003}).

\bibitem{simmonsbeadstiff96}
\bibinfo{author}{Simmons, R.~M.}, \bibinfo{author}{Finer, J.~T.},
  \bibinfo{author}{Chu, S.} \& \bibinfo{author}{Spudich, J.~A.}
\newblock \bibinfo{title}{Quantitative measurements of force and displacement
  using an optical trap}.
\newblock \emph{\bibinfo{journal}{Biophysical Journal}}
  \textbf{\bibinfo{volume}{70}}, \bibinfo{pages}{1813} (\bibinfo{year}{1996}).

\bibitem{gregory2005bayesian}
\bibinfo{author}{Gregory, P.}
\newblock \emph{\bibinfo{title}{Bayesian Logical Data Analysis for the Physical
  Sciences: A Comparative Approach with Mathematica{\textregistered} Support}}
  (\bibinfo{publisher}{Cambridge University Press}, \bibinfo{year}{2005}).
\end{thebibliography}
\end{document}